\documentclass[showpacs,floatfix,twocolumn,aps,nofootinbib]{revtex4-1}

\usepackage{graphicx}
\usepackage{epstopdf}
\usepackage{latexsym}
\usepackage{amsmath}
\usepackage{amssymb}
\usepackage{color}
\usepackage{mathrsfs}

\usepackage[center]{subfigure}

\begin{document}

 \newcommand{\bq}{\begin{equation}}
 \newcommand{\eq}{\end{equation}}
 \newcommand{\bqn}{\begin{eqnarray}}
 \newcommand{\eqn}{\end{eqnarray}}
 \newcommand{\nb}{\nonumber}
 \newcommand{\lb}{\label}
 \newcommand{\be}{\begin{equation}}
\newcommand{\en}{\end{equation}}
\newcommand{\PRL}{Phys. Rev. Lett.}
\newcommand{\PL}{Phys. Lett.}
\newcommand{\PR}{Phys. Rev.}
\newcommand{\CQG}{Class. Quantum Grav.}

\title{A Charged Compact Star Solution for a Conducting Sphere}
\author{Kai Lin}\email{lk314159@hotmail.com}
\affiliation{Departamento de Física, Centro de Ciências Exatas e da Natureza, Universidade Federal da Paraíba, CEP 58059-970, João Pessoa, Brazil}
\affiliation{Universidade Estadual da Paraíba, Rua Baraúnas, 351, Bairro Universitário, Campina Grande, Brazil}
\affiliation{Universidade Federal de Campina Grande, Campina Grande, PB 58429-900, Brasil}
%


\begin{abstract}
This paper considers the simplest charged stellar model, namely, a uniform density star with electric charge distributed on its surface. In this model, we treat the uniform density star as a conducting sphere, where all electric charges are uniformly distributed on the surface and represented by a delta function. We obtain an analytical solution for the background metric of the star spacetime and find that the pressure of the perfect fluid inside the star is directed inward (negative pressure) near the inner surface. This is because the repulsive force between surface electric charges requires an inward pressure to maintain equilibrium. However, when the charge becomes too large, the pressure of the ideal fluid at the stellar center gets directed outwards again. This is because the electric field itself is also a part of the stress–energy tensor $T_{\mu\nu}$, acting as a source of gravity. When the gravitational effect of the electric field dominates, the perfect fluid must provide outward pressure to balance the inward gravitational attraction.On the other hand, we have also studied the axial perturbation in charged spacetimes using the finite difference method. We found that axial Maxwell perturbation and axial gravitational perturbation are coupled in the star spacetime, making it difficult to separate the variables by coordinate transformation. The computational results show that the waveforms of quasinormal modes vary with different charges.
\end{abstract}


\maketitle

\section{Introduction}
\renewcommand{\theequation}{1.\arabic{equation}} \setcounter{equation}{0}
Since LIGO detected the first gravitational wave event from a binary black hole merger\cite{LIGO1}, astronomy has entered the era of gravitational waves\cite{LIGO2,LIGO3,LIGO4,LIGO5}, making massive compact stars a central topic in modern astrophysical research. Shortly thereafter, the LIGO-VIRGO collaboration detected gravitational waves from a neutron star merger\cite{LIGO6,LIGO7}, marking the first time humanity observed the same astronomical event using both gravitational and electromagnetic waves—two fundamentally different physical messengers. This milestone ushered in the era of multi-messenger astronomy. The study of neutron stars is of great significance, as their properties provide a testing ground for both particle physics and gravitational theories. Moreover, the precise measurements of neutron star characteristics enable applications such as galactic positioning, neutron star navigation, and even gravitational wave detection.

At cosmological scales, gravity is the dominant interaction, but electromagnetic interactions can also play a crucial role in certain scenarios. For instance, neutron stars can possess extremely strong magnetic fields, forming magnetars \cite{magnetar1,magnetar2}. In black hole physics, according to the no-hair theorem, a black hole is characterized externally by only three parameters: mass, charge, and angular momentum. What's more, for a given mass and angular momentum, the charge of a black hole cannot be too large; otherwise, the event horizon would vanish, leading to a naked singularity, which is forbidden by the cosmic censorship conjecture. Investigating the relationship between electromagnetic and gravitational fields can provide valuable insights and criteria for modified gravity theories and advancing quantum gravity research.

On the other hand, in recent years, a lot of activities are directed to the observation of the so-called Fast Radio Bursts \cite{FRBs}, which present a new challenge in the understanding of astrophysical phenomena \cite{Petroff,FRBEnigma}.
Moreover, it has been suggested that magnetars and solutions with strong magnetic fields may play a definite role in Fast Radio Bursts \cite{bocheneketal}. Observations are from relatively small frequencies (about 100 MHz) \cite{Pleunis2021,Pastor2021} to 8 GHz \cite{Gajjar2018}. 

We thus tried to find whether instabilities can occur in solutions with magnetic fields. They might get associated to instabilities generating bursts. Recently we have considered some nontrivial solutions with magnetic fields, but no instability has been found, in spite of the complexity of the solution, to render perturbations unstable \cite{eveling2025,Marcos2026}.

As a matter of fact, several solutions that might display instabilities, such as time-dependent solutions, have been found to be stable under perturbations \cite{shaowangabdalla,abdallachirentisaa01,abdallachirentisaa02}. One notable exception was the case of charged black holes \cite{zhangwangabdalla,zhangpanwangabdalla,zhuzhangpellicerwangabdalla}.

There is an observational chance to grasp details of FRB's in a new generation of radio telescopes (see e.g. \cite{abdallaetal,FRBabdus}). 

Based on the above motivations, this paper investigates a charged uniform density star and its axial electromagnetic and gravitational wave perturbations. We consider the simplest stellar model, assuming that the star is composed of a perfect fluid with a constant density. Furthermore, we assume that the perfect fluid constituting the star is a good conductor, making the star effectively a conducting sphere. Due to charge screening effects, all charge is distributed on the surface of this conducting sphere. As a result, the exterior spacetime of the star is described by the Reissner–Nordstr\"om metric, while the electric potential inside the star remains constant. In section 2, we derive the analytical solution for the metric of this spacetime. The fact that all charge resides on the stellar surface leads to a discontinuity in the pressure of the perfect fluid at the surface. In section 3, we derive the axial perturbation equation, but it is found that it is hard to obtain decoupling equations. In Section 4, we analyze axial perturbations of this spacetime using the finite difference method. Finally, we also discuss the properties of the stellar radius as the system approaches the extremal black hole limit. At this stage, the pressure inside the star is entirely outward-directed, so that during gravitational collapse the charged matter is expelled due to electromagnetic interactions, preventing the formation of an extremal black hole or a naked singularity.. Section 5 presents further discussions and conclusions.

\section{background metric of charged uniform density star}
\renewcommand{\theequation}{2.\arabic{equation}} \setcounter{equation}{0}

In curved spacetime the action of the Einstein-Maxwell Equation with source is given by
 \bqn
 \lb{action1}
S&=&\frac{1}{2\kappa}\int\sqrt{-g}d^4x\left[R-2\Lambda-\kappa{\cal L}_m\right]\nb\\
&&-\frac{1}{4\mu_0}\int\sqrt{-g}dx^4F_{\mu\nu}F^{\mu\nu}-\int dx^4 A_\mu J^\mu~~~~
 \eqn
where $J^\mu\equiv ceu^\mu_J$ is 4-current density in curved spacetime, $\kappa\equiv8\pi G/c^4$, $c^2=1/\mu_0\epsilon_0$, ${\cal L}_m$ is the Lagrange term of perfect fluid matter, $ceA_\mu u^\mu_J$ is the source term, $u^\mu_J$ the 4-velocity of electric charge $e$. It satisfies the normalization condition $g_{\mu\nu}u^\mu_J u^\nu_J=-1$. $F_{\mu\nu}\equiv \nabla_\mu A_\nu-\nabla_\nu A_\mu$ is the electromagnetic tensor and $A_\mu$ the corresponding four-potential.

According to the above action, we choose $(g^{\mu\nu},A_\mu,J^\mu)$ as the basic variables, so that the Einstein-Maxwell equations are given by
\bqn
 \lb{action2}
&&R_{\mu\nu}-\frac{1}{2}g_{\mu\nu}R+g_{\mu\nu}\Lambda
=\kappa\left(T^m_{\mu\nu}+T^{EM}_{\mu\nu}\right)\nb\\
&&T^{m}_{\mu\nu}\equiv \left(p+\rho c^2\right)u_\mu u_\nu+pg_{\mu\nu}\nb\\
&&T^{EM}_{\mu\nu}\equiv \frac{1}{\mu_0}\left(F_{\mu\alpha}F_\nu^\alpha-\frac{1}{4}g_{\mu\nu}F_{\alpha\beta}F^{\alpha\beta}\right)\nb\\
&&\frac{\partial}{\partial x^\nu}\left(\sqrt{-g}F^{\mu\nu}\right)=\mu_0J^\mu\nb\\
&&\nabla^\mu\left(T^m_{\mu\nu}+T^{EM}_{\mu\nu}\right)=0
 \eqn

It is worth noting that although both $J^\mu$ and $T^m_{\mu\nu}$ contain the four-velocity in their definitions, the physical meanings of 
the 4-velocities in the two cases are essentially different. In the former, $u^\mu_J$ represents the four-velocity of electric charge $e$, whereas in the latter $u_\mu$ (which also satisfies the relation $g_{\mu\nu}u^\mu u^\nu=-1$) it denotes the four-velocity of the fluid matter with $\rho$ and $p$. In fact, $J^\mu$ should be regarded as an external source, and by specifying its form one can obtain different spacetime metrics.

The metric of a statically charged uniform density star is given by
 \bqn
 \lb{Metric1}
ds^2&=&-f(r)c^2dt^2+\frac{dr^2}{h(r)}+r^2d\theta^2+r^2\sin^2\theta d\varphi^2,\nb\\
A_\mu &=& A_t(r)\delta_\mu^t,\nb\\
u^\mu&=&\frac{\delta^\mu_t}{c\sqrt{f(r)}}.
 \eqn
In the model of a uniform density star, $\rho=\rho_c$ is a constant inside the star, while outside the star's surface $r_s$, $\rho = 0$, i.e.
 \bqn
 \lb{Metric2}
\rho=\rho_c[1-U(r-r_s)]
 \eqn
where $U$ is the unit step function
 \bqn
 \lb{Metric4}
U(r)=\left\{
\begin{array}{cc}
    0 & r<0\\
    1  & r>0 \\
\end{array}
  \right.
 \eqn
If we consider the uniform density star as a conductor, due to the electrostatic shielding effect, all the electrical charges should be uniformly distributed on the surface $r_s$ of the sphere. Thus, we can set
 \bqn
 \lb{Metric3}
e=\sigma\delta(r-r_s)
 \eqn
where the Dirac delta function $\delta(r)\equiv dU(r)/dr$ and $\sigma$ are parameters depending on the electric charge.

From Einstein field equation and Maxwell equation, we can get 4 independent equations
 \bqn
 \lb{FieldEqu1}
f'&=&-\frac{f}{r}+\frac{rf}{h}\left(\frac{1}{r^2}-\Lambda+\kappa p\right)-\frac{\kappa}{2\mu_0}rA_t'^2,\nb\\
h'&=&\frac{1}{r}-\Lambda r-\kappa c^2 r\rho-\left(\frac{1}{r}+\frac{\kappa}{\mu_0}\frac{r}{2f}A_t'^2\right)h,\nb\\
p'&=&\left[\frac{h-1+\Lambda r^2}{2rh}-\kappa\frac{r}{2h}p+\frac{\kappa}{\mu_0}\frac{r}{4f}A_t'^2\right](c^2\rho+p)\nb\\
&&-c\frac{A_t'}{\sqrt{f}}\sigma\delta(r-r_s),\nb\\
A_t''&=&\kappa\frac{r}{2h}(c^2\rho+p)A_t'-\frac{2}{r}A_t'-\mu_0c\frac{\sqrt{f}}{h}\sigma\delta(r-r_s)\nb\\
 \eqn

Let's calculate the junction conditions at the surface $r_s$ of the star. In spacetime of charged uniform density star, we can reasonable assume the results of $f$, $h$, $p$ and $A_t'$ are
 \bqn
 \lb{FieldEquJunction1}
f(r)&=&f_i(r)[1-U(r-r_s)]+f_o(r)U(r-r_s)\nb\\
h(r)&=&h_i(r)[1-U(r-r_s)]+h_o(r)U(r-r_s)\nb\\
p(r)&=&p_i(r)[1-U(r-r_s)]+p_o(r)U(r-r_s)\nb\\
A_t'(r)&=&A_{ti}'(r)[1-U(r-r_s)]+A_{to}'(r)U(r-r_s)\nb\\
 \eqn
Considering the relation
 \bqn
 \lb{FieldEquJunction2}
&&\lim_{\epsilon\to 0}\int_{r_s-\epsilon}^{r_s+\epsilon}{\cal F}(r,U(r-r_s))dr=0,\nb\\
&&~~~\text{where}~{\cal F}(r,U)~\text{remains finite on}~[r_s-\epsilon,r_s-\epsilon]\nb\\
&&\lim_{\epsilon\to 0}\int_{r_s-\epsilon}^{r_s+\epsilon}\delta(r-r_s)dr=1\nb\\
&&\lim_{\epsilon\to 0}\int_{r_s-\epsilon}^{r_s+\epsilon}\frac{\partial F}{\partial U}\delta(r-r_s)dr=F(r_s,U)|^{U=1}_{U=0},
 \eqn
and combining with Eq.(\ref{Metric2}), we integrate Eq.(\ref{FieldEqu1}) from $r_s-\epsilon$ to $r_s+\epsilon$ with $\epsilon\rightarrow0$ to obtain:
 \bqn
 \lb{FieldEquJunction3}
f(r_s+0)&=&f(r_s-0)=f(r_s)\nb\\
h(r_s+0)&=&h(r_s-0)=h(r_s)\nb\\
p(r_s+0)&=&p(r_s-0)-\frac{A_t'(r_s+0)+A_t'(r_s-0)}{2\sqrt{f(r_s)}}c\sigma\nb\\
A'(r_s+0)&=&A'(r_s-0)-\mu_0\frac{\sqrt{f(r_s)}}{h(r_s)}c\sigma.
 \eqn
What's more, from the relation $dA_t(r)=A_t'(r)dr$, we have
 \bqn
 \lb{FieldEquJunction4}
A_t(r_s+0)-A_t(r_s-0)=\lim_{\epsilon\to 0}\int_{r_s-\epsilon}^{r_s+\epsilon}A_t'(r)dr&&\nb\\
=\lim_{\epsilon\to 0}\int_{r_s-\epsilon}^{r_s+\epsilon}\left\{A_{ti}'(r)[1-U(r-r_s)]\right.&&\nb\\
\left.+A_{to}'(r)U(r-r_s)\right\}dr&=&0,\nb\\
 \eqn
or
 \bqn
 \lb{FieldEquJunction5}
A_t(r_s+0)=A_t(r_s-0)=A_t(r_s).
 \eqn
The field equations above (\ref{FieldEqu1}), together with the junction conditions (\ref{FieldEquJunction3})-(\ref{FieldEquJunction5}) and the equation of state (\ref{Metric2}) for an charged uniform star, constitute a complete set of equations for solving the spacetime geometry.

Outside of the star $r>r_s$, it is vacuum so $p=0$, and it is no other than the Reissner–Nordstr\"om solution:
 \bqn
 \lb{FieldEqu2}
f_o(r)=h_o(r)&=&1-\frac{2GM}{c^2r}+\frac{GQ^2}{4\pi\epsilon_0c^4r^2}-\frac{\Lambda}{3}r^2\nb\\
A_{to}(r)&=&\frac{Q}{4\pi\epsilon_0cr}
 \eqn

On the other hand, inside of the star $r<r_s$, since the charge is distributed all over the surface of the star, so $A_t$ becomes a constant and the solution is given by:
 \bqn
 \lb{FieldEqu3}
h_i(r)&=&1-\frac{\kappa}{3}\rho_cc^2r^2-\frac{\Lambda}{3}r^2\nb\\
f_i(r)&=&f_C\frac{\left[C_{\cal A}\sqrt{h_i(r)}-3\kappa C_{\cal B}\sqrt{h_i(r_s)}\right]^2}{4(\Lambda+\kappa\rho_cc^2)^2}\nb\\
p_i(r)&=&\frac{\rho_cc^2C_{\cal A}\sqrt{h_i(r)}-(\kappa\rho_cc^2-2\Lambda)C_{\cal B}\sqrt{h_i(r_s)}}{-C_{\cal A}\sqrt{h_i(r)}+3\kappa C_{\cal B}\sqrt{h_i(r_s)}}\nb\\
C_{\cal A}&\equiv&3\kappa P_s+\kappa\rho_cc^2-2\Lambda\nb\\
C_{\cal B}&\equiv&P_s+\rho_cc^2
 \eqn
where $p_i(r_s)\equiv p(r_s-0)\equiv P_{S}$, and in full spacetime, $p$ can be rewritten as
 \bqn
 \lb{FieldEqu4}
p(r)&=&p_i(r)[1-U(r-r_s)]
 \eqn
and we can set $A_t$ to be continuous at the surface of the star, the constant $A_t$ being given by
 \bq
 \lb{FieldEqu6}
A_{ti}(r)\equiv A_t(r_s-0)=A_t(r_s+0)=\frac{Q}{4\pi\epsilon_0cr_s}\equiv A_s~~~ \quad .
 \eq
For $A_t'(r)$ in full spacetime we thus have
 \bqn
 \lb{FieldEqu7}
A_t'(r)&=&-\frac{Q}{4\pi\epsilon_0cr^2}U(r-r_s)
 \eqn
 


According to the junction condition (\ref{FieldEquJunction3}) at surface $r=r_s$, 
 \bqn
 \lb{Solution1}
\rho_c&=&\frac{3M}{4\pi r_s^3}\left(1-\frac{\mu_0Q^2}{8\pi M r_s}\right),\nb\\
P_s&=&-\frac{\mu_0c^2Q^2}{32\pi^2r_s^4},\nb\\
\sigma&=&\frac{Q}{4\pi r_s^2}\sqrt{1-\frac{\Lambda}{3}r_s^2+G\frac{\mu_0Q^2-8\pi Mr_s}{4\pi c^2r_s^2}},\nb\\
f_C&=&1.
 \eqn
From the above equation, we can observe several interesting conclusions. Since the terms containing 
$\mu_0$ and 
$G$ represent the effects of the electromagnetic field and the gravitational field respectively, while $\Lambda$ also corresponds to a gravitational effect, we can see that the density $\rho_c$ of the star is not only related to its total mass but is also influenced by the electromagnetic field. Moreover, under the influence of the electromagnetic field, the pressure $P_s$ at the inner surface of the star becomes negative. This occurs because like charges repel each other, and the fluid must provide an attractive effect at this location to counterbalance the repulsive force. On the other hand, the surface charge density $\sigma$ of the star is also affected by gravitational effects, leading to a correction term.

In order to simplify the calculation, we focus only on the case of $\Lambda = 0$, and introduce the dimensionless quantities as follows
\bqn
 \lb{dimensionless1}
M_o&=&\frac{G}{c^2r_s}M,\nb\\
Q_o&=&\frac{\sqrt{G}}{2c^2r_s\sqrt{\epsilon_0\pi}}Q,\nb\\
\rho_o&=&r_s^2c^2\kappa\rho,\nb\\
P_o&=&r_s^2\kappa P_s,\nb\\
x&=&\frac{r}{r_s},\nb\\
\tau&=&\frac{ct}{r_s},\nb\\
w&=&\frac{r_s\omega}{c},
\eqn
and the function transformation
\bqn
 \lb{dimensionless2}
f_x(x)&=&f(r),\nb\\
h_x(x)&=&h(r),\nb\\
A_{tx}(x)&=&\sqrt{\frac{\kappa}{\mu_0}}A_t(r),\nb\\
p_x(x)&=&\kappa r_s^2p(r),\nb\\
\rho_x(x)&=&c^2\kappa r_s^2\rho(r),
\eqn
so that the dimensionless function solution are given by
\bqn
 \lb{dimensionless4}
f_x(x)&=&f_{xi}(x)[1-U(x-1)]+f_{xo}(x)U(x-1),\nb\\
h_x(x)&=&h_{xi}(x)[1-U(x-1)]+h_{xo}(x)U(x-1),\nb\\
A_{tx}(x)&=&\sqrt{2}Q_o[1-U(x-1)]+\frac{\sqrt{2}Q_o}{x}U(x-1),\nb\\
p_x(x)&=&p_{xi}(x)[1-U(x-1)],\nb\\
\rho_x(x)&=&\rho_o[1-U(x-1)],
 \eqn
where 
\bqn
 \lb{dimensionless5}
f_{xo}(x)&=&h_{xo}(x)=1-\frac{2M_o}{x}+\frac{Q_o^2}{x^2},\nb\\
h_{xi}(x)&=&1-2M_ox^2+Q_o^2x^2,\nb\\
f_{xi}(x)&=&\left[\frac{(3M_o-2Q_o^2)\sqrt{h_{xi}(1)}-(M_o-Q_o^2)\sqrt{h_{xi}(x)}}{2M_o-Q_o^2}\right]^2,\nb\\
p_{xi}(x)&=&(Q_o^2-2M_o)\left[(3M_o-2Q_o^2)\sqrt{h_{xi}(1)}\right.\nb\\
&&\left.-3(M_o-Q_o^2)\sqrt{h_{xi}(x)}\right]\left[(3M_o\right.\nb\\
&&\left.-2Q_o^2)\sqrt{h_{xi}(1)}-(M_o-Q_o^2)\sqrt{h_{xi}(x)}\right]^{-1},\nb\\
\rho_o&=&6M_o-3Q_o^2.
 \eqn
The continuity condition $f_{xi}(1)=f_{xo}(1)$, $h_{xi}(1)=h_{xo}(1)$ and $p_{xi}(1)\equiv P_{S}=-Q_o^2$, and 
\bqn
 \lb{dimensionless4a}
\sigma_o&=&{2\sqrt{\pi G\mu_0}r_s}\frac{\sigma}{c}=Q_o\sqrt{1-2M_o+Q_o^2}.
\eqn

In this paper, the results of the analytical solution with $\Lambda=0$ for different parameter values are plotted in FIG.\ref{metricFig}.
\begin{figure*}[!htbp]
\centering
\includegraphics[width=0.7\columnwidth]{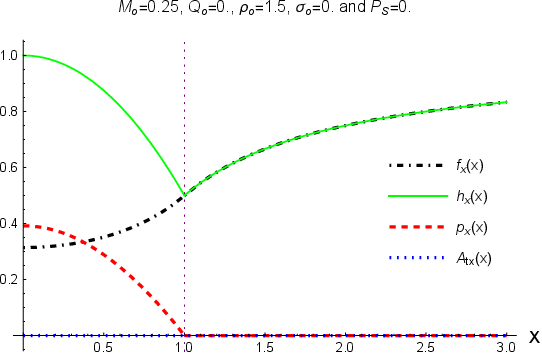}\includegraphics[width=0.7\columnwidth]{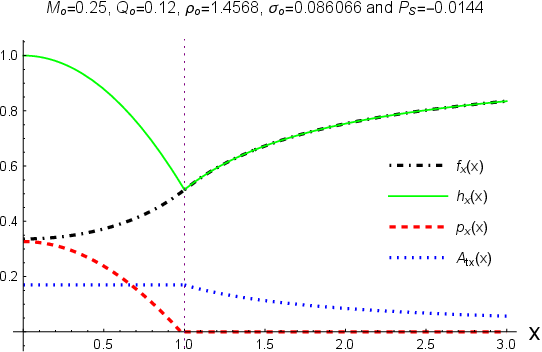}\includegraphics[width=0.7\columnwidth]{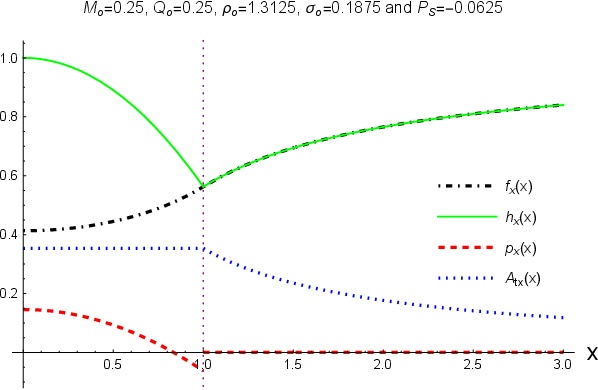}
\includegraphics[width=0.7\columnwidth]{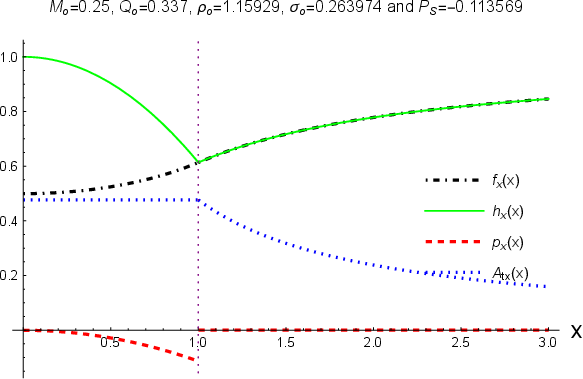}\includegraphics[width=0.7\columnwidth]{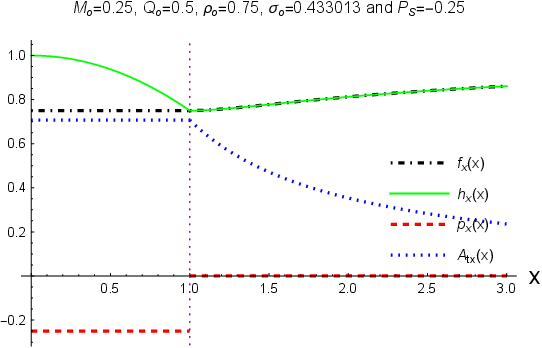}\includegraphics[width=0.7\columnwidth]{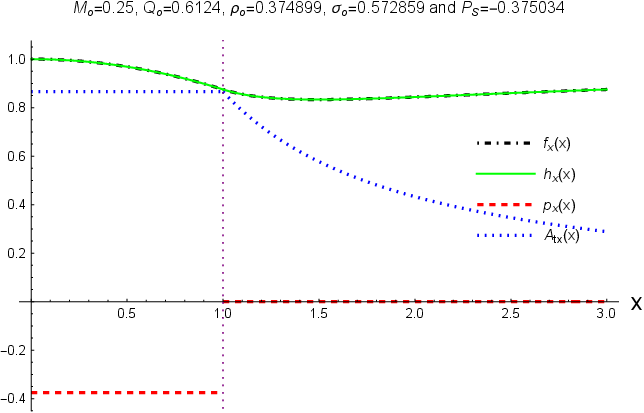}
\includegraphics[width=0.7\columnwidth]{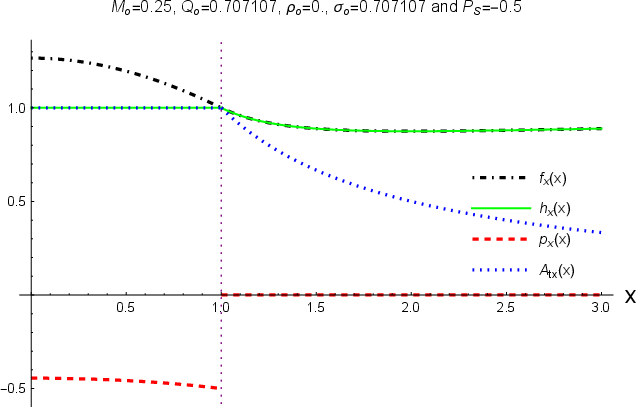}\includegraphics[width=0.7\columnwidth]{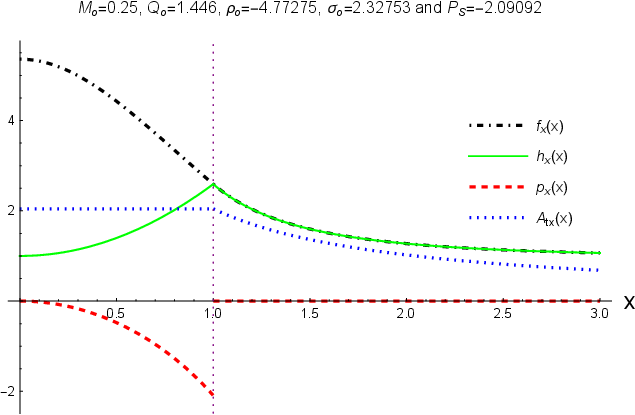}\includegraphics[width=0.7\columnwidth]{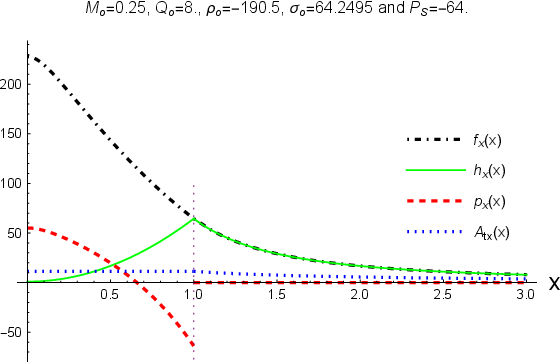}
\caption{$h_x(x)$, $f_x(x)$, $p_x(x)$ and $A_x(x)$ with $M_o=1/4$. We find that for $0.337<|Q_o|<1.446$, the pressure inside the star is entirely negative. Outside this range, the pressure becomes negative only near the stellar surface, while it remains positive in the central region. The negative-pressure region inside the star requires the stellar matter to provide an attractive effect in order to balance the repulsive force induced by the electrostatic interaction.}
\lb{metricFig}
\end{figure*}


From FIG.\ref{metricFig}, it can be observed that, for a fixed $M_o$, as $Q_o$ increases, the pressure $P_S$ provided by the perfect fluid at the inner surface of the star gradually decreases. This is due to the mutual repulsive effect of like-signed electrostatic charges on the star’s surface, necessitating an inward positive pressure at the inner surface to balance this effect. However, when $Q_o$ reaches the critical value $Q_m$, the pressure at the inner surface of the star begins to weaken and may even become negative. This arises because the charges and electromagnetic fields not only produce electromagnetic effects but also contribute to the stress–energy tensor $T_{\mu\nu}$, thereby inducing gravitational effects. When the surface charge of the star is sufficiently large, the stress–energy tensor of the electric field dominates, and the resulting gravitational effect is strong enough to counteract the electrostatic repulsive force. Consequently, when the charge is sufficiently large, the pressure of the perfect fluid inside the star reverses, becoming an outward repulsive force. It is worth noting, as $M_o<Q_o^2/2$, a normal perfect fluid with $\rho_o > 0$ cannot provide the corresponding negative pressure. Therefore, the perfect fluid in this case must be an exotic fluid with $\rho_o < 0$.

For $M_o=Q_o^2/2$, we have $\rho=0$, and the metric of star is
\bqn
 \lb{especial1}
f_{xi}(x)&=&\frac{\left[Q_o^2(x^2-1)-4\right]^2}{16},\nb\\
h_{xi}(x)&=&1,\nb\\
h_{xo}(x)&=&f_{xo}(x)=1-\frac{Q_o^2}{x}+\frac{Q_o^2}{x^2},\nb\\
p_{xi}(x)&=&\frac{4Q_o^2}{Q_o^2(x^2-1)-4},\nb\\
\rho_{o}(x)&=&0.
 \eqn

On the other hand, we also identify several two cases.

For $M_o=Q_o^2$, $\frac{dp_{xi}(x)}{dx}=0$, so the pressure inside the star is a constant, and the metric of star is given by
\bqn
 \lb{especial2}
f_{xi}(x)&=&1-Q_o^2,\nb\\
h_{xi}(x)&=&1-Q_o^2x^2,\nb\\
h_{xo}(x)&=&f_{xo}(x)=1-2\frac{Q_o^2}{x}+\frac{Q_o^2}{x^2},\nb\\
p_{xi}(x)&=&-Q_o^2,\nb\\
\rho_{o}(x)&=&3Q_o^2.
 \eqn

For $M_o=2Q_o^2/3$, the function $f_x(x)$ and $h_x(x)$ coincide, and the pressure inside the star is constant. The metric is
\bqn
 \lb{especial3}
f_{xi}(x)&=&h_{xi}(x)=1-\frac{Q_o^2x^2}{3},\nb\\
h_{xo}(x)&=&f_{xo}(x)=1-\frac{4}{3}\frac{Q_o^2}{x}+\frac{Q_o^2}{x^2},\nb\\
p_{xi}(x)&=&-Q_o^2,\nb\\
\rho_{o}(x)&=&Q_o^2.
 \eqn

Finally, we discuss the intrinsic properties of the charged star itself. From the above analysis, we find that if the existence of exotic matter is allowed, then for a fixed stellar mass, the total electric charge of the star can, in principle, be arbitrarily large. However, we also know that for black holes the electric charge cannot exceed the black hole mass. Therefore, it is necessary to investigate how the metric coefficients of the star behave as the system approaches the extremal black hole limit.

\begin{figure*}[!htbp]
\centering
\includegraphics[width=1\columnwidth]{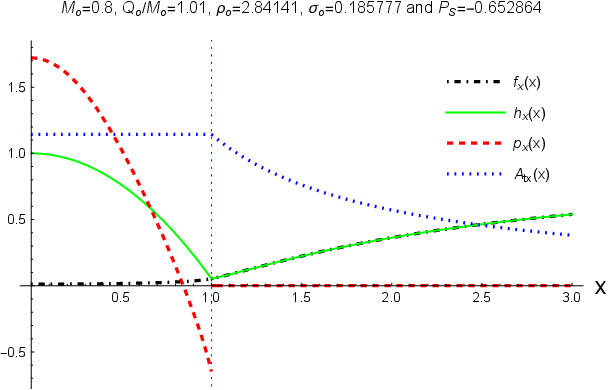}\includegraphics[width=1\columnwidth]{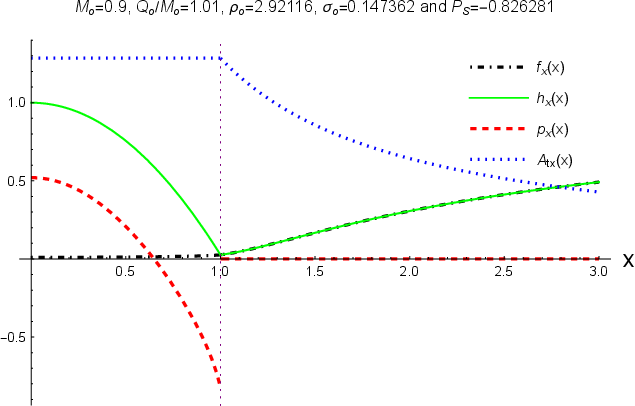}
\includegraphics[width=1\columnwidth]{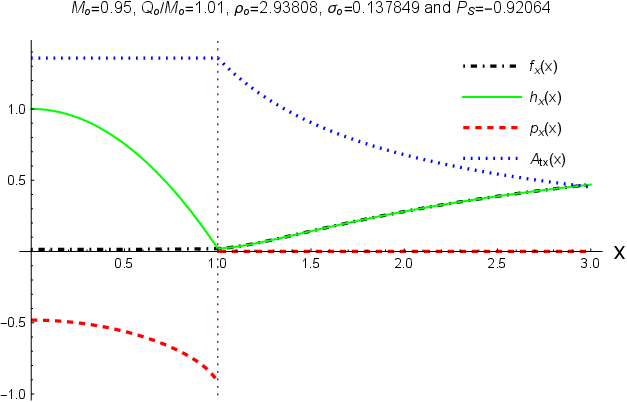}\includegraphics[width=1\columnwidth]{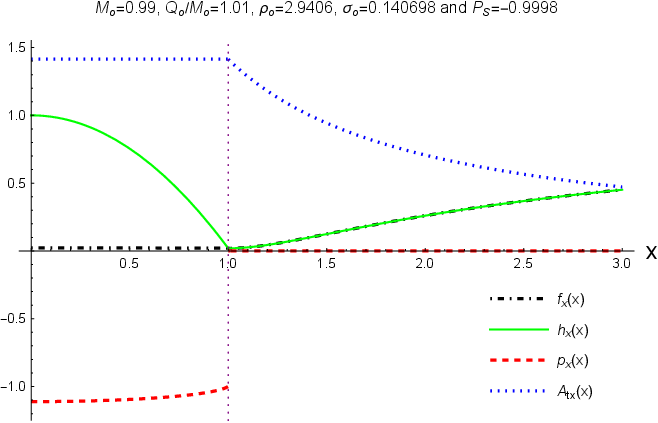}
\caption{$h_x(x)$, $f_x(x)$, $p_x(x)$, $A_x(x)$ with $Q_o/M_o=1.01$}
\lb{MExtFig}
\end{figure*}

In Fig. \ref{MExtFig}, we fix $Q_o/M_o=1.01$, i.e., the total charge is slightly larger than the mass, and then compare cases in which $M_o$ gradually approaches the critical value for gravitational collapse. It is straightforward to see that, as the stellar radius approaches the black hole event horizon, although the stellar matter remains in an ordinary state, the internal pressure of such a star with charge slightly exceeding its mass becomes entirely negative. In other words, there is no longer an inward pressure inside the star; instead, a purely repulsive effect emerges. This repulsion is clearly provided by the electrostatic interaction among the charges, which prevents the star from undergoing further collapse.

In a realistic collapse process, charged matter can separate from neutral matter. The neutral component continues to collapse under gravity, while the charged component experiences an outward repulsive force during the collapse and is consequently expelled. As a result, the black hole formed at the end of the collapse cannot carry a charge larger than its mass, and therefore neither an extremal black hole nor a naked singularity can be formed.

In the next section, we derive the axial perturbation equation in this spacetime background.

\section{axial perturbation equations}
\renewcommand{\theequation}{3.\arabic{equation}} \setcounter{equation}{0}

We know that both gravitational waves in general relativity and electromagnetic waves in Maxwell’s theory have two polarization directions. In the study of quasinormal modes, perturbations are usually classified into axial and polar perturbations based on the parity of the perturbation quantities. This paper focuses on the relatively simpler axial perturbations. Under axial perturbations, the perturbation equations can often be simplified into a very concise form. In static black hole spacetimes or constant-density star spacetimes without charge, the axial perturbation equations can even be decoupled into a second-order differential equation. Moreover, since the radial equation does not depend on the magnetic quantum number $m$, we can choose the Regge-Wheeler gauge\cite{Regge} as follows:
\bqn
\lb{perturbation1}
\delta g_{\mu\nu}&=&
\begin{bmatrix}
0 & 0 & 0 & h_0(r) \\
0 & 0 & 0 & h_1(r) \\
0 & 0 & 0 & 0  \\
h_0(r) & h_1(r) & 0 & 0
\end{bmatrix}
e^{-i\omega t}\sin\theta\partial_\theta P_L(\cos\theta)\nb\\
\delta A_\mu&=&\sqrt{\frac{\kappa}{\mu_0}}\delta^\varphi_\mu \frac{i}{\omega}h_A(r)e^{-i\omega t}\sin\theta\partial_\theta P_L(\cos\theta)\nb\\
\delta u_\mu&=&\delta^\varphi_\mu h_n(r)e^{-i\omega t}\sin\theta\partial_\theta P_L(\cos\theta)\nb\\
\delta u_J&=&\delta^\varphi_\mu h_J(r)e^{-i\omega t}\sin\theta\partial_\theta P_L(\cos\theta)
\eqn
where $P_L(\cos\theta)$ is Legendre function, which is proportional to the spherical function $Y_{l,m=0}(\theta,\varphi)$.

It is also worth emphasizing once again that 
$J^\mu$ (or $u^\mu_J$) is an external source, and its perturbation plays the role of a driving force. Therefore, one may choose $h_J$ in an artificial manner. In principle, the perturbation of $J^\mu$ can be specified quite arbitrarily. However, in this work we restrict ourselves to axial perturbations only. Other types of perturbations of $u^\mu_J$ may excite modes beyond the axial sector and are therefore outside the scope of the present study.

Substituting Eq.(\ref{perturbation1}) into Einstein field equation and Maxwell field equation, we find 
\bqn
\lb{perturbation1a}
h_n=-\frac{h_0(r)}{cr^2\sqrt{f(r)}}+ic\sigma \frac{h_A(r)}{r^2\omega(p+c^2\rho)}\delta(r-r_s). 
\eqn

However, we find that even under the simplest axial perturbations, it is still impossible to obtain a fully decoupled Schrödinger equation as the master equation for quasinormal modes. Instead, we derive a system of two coupled second-order differential equations. By using the dimensionless quantities in Eq.(\ref{dimensionless1}) and Eq.(\ref{dimensionless2}), the coupled axial perturbation equations are given by
\bqn
 \lb{perturbation2}
\partial_y^2R_T(x)+\left(w^2-V_T(x)\right)R_T(x)&=&U_T(x)R_V(x)+F(x),\nb\\
\partial_y^2R_V(x)+\left(w^2-V_V(x)\right)R_V(x)&=&U_V(x)R_T(x),
\eqn
where $y=\int_0^x\frac{dx}{\sqrt{f_x(x)h_x(x)}}$, $h_1=B_x(x)R_T(x)$, $h_A=R_V(x)$, the potentials are given by
\bqn
 \lb{perturbation3}
V_T(x)&=&\frac{f_x}{2x^2}\left[6h_x-x^2(p_x-\rho_x)+2L(L+1)-6\right]\nb\\
&&+\frac{1}{2}h_xA_x'^2\nb\\
U_T(x)&=&-2\frac{A_x'}{B_x},\nb\\
V_V(x)&=&2h_xA_x'^2+L(L+1)\frac{f_x}{x^2},\nb\\
U_V(x)&=&-(L^2+L-2)\frac{B_xf_xh_xA_x'}{x^2},\nb\\
F(x)&=&i\frac{\omega}{c^2}\sqrt{\frac{8\pi G}{\mu_0}}\sigma_er_s^2\mu_0\sqrt{f_x}\nb\\
&&\times(h_0+cr^2\sqrt{f_x}h_J) \delta_x(x-1),
\eqn
and $B_x$ is
\bqn
 \lb{perturbation4}
B_x(x)&=&B_{xi}(x)[1-U(x-1)]+B_{xo}(x)U(x-1),\nb\\
B_{xo}&=&\frac{\mu_0}{4\pi c}\frac{B_wx^3}{Q_o^2+x(x-2M_o)},\nb\\
B_{xi}&=&\frac{\mu_0}{4\pi c}\frac{B_w\sqrt{2M_o-Q_o^2}x}{\sqrt{1-2M_ox^2+Q_ox^2}}\nb\\
&&\times\left[(M_o-Q_o^2)^2-4Q_o^4+12M_oQ_o^2\right.\nb\\
&&\left.-3Q_o^2-9M_o^2+4M_o\right]^{-1/2}.
\eqn
where we choose $B_w=-\frac{4\sqrt{2}c\pi}{\mu_0\sqrt{L^2+L-2}}$ so that $U_T(x)=U_A(x)$. On the other hand $F(x)$ term represents the influence of external sources on the system, but here we mainly study the natural vibration (or free vibration) of the system itself, so we can choose $h_J = -h_0/cr^2\sqrt{f_x}$ to make this term disappear.

Next, we discuss the boundary conditions that $R_T$ and $R_V$ must satisfy at the surface of the star. Without loss of generality, we can first assume at the stellar surface:
\bqn
\lb{perturbation6}
R_T(1-0)&=&R_T(1+0)\equiv R_T(1)\nb\\
R_V(1-0)&=&R_V(1+0)\equiv R_V(1)
\eqn
Then let's put $\int_{1-0}^{1+0}dx$ both side of the perturbation equations (\ref{perturbation2}) to integrate, and we obtain the boundary condition
\bqn
\lb{perturbation7}
&&R_T'(1+0)-R_T'(1-0)=0\nb\\
&&R_V'(1+0)-R_V'(1-0)=0
\eqn
This relationship indicates that $R_V$ and $R_T$ are continuous at the stellar surface, but $R_T$ is not smooth here. Since the parameter $\Theta$ is proportional to $Q$, this lack of smoothness is entirely caused by the charge distributed on the stellar surface. 

In the next section, we will use the finite difference method to compute the quasinormal mode oscillations described by Eq. (\ref{perturbation4}).

\section{numerical calculation by using finite difference method}
\renewcommand{\theequation}{4.\arabic{equation}} \setcounter{equation}{0}

Applying an axial perturbation to a compact star, its dynamical behavior can be described by the perturbation equations derived in the previous section. For a compact star (but not a black hole), the wave equations governing this perturbation must satisfy the boundary conditions: the perturbation remains finite at the center of the star and consists solely of outgoing waves at infinity.  

Due to the presence of a potential function in the perturbation equation, analytical solutions are usually difficult to obtain. Various numerical methods have been developed to solve quasinormal mode equations in different spacetimes\cite{QNM1,QNM2,QNM3,QNM4,QNM5,QNM6}, but most of these methods are designed for perturbations in black hole spacetimes, whereas numerical approaches for perturbations in compact stars remain relatively limited. Nevertheless, the finite difference method (FDM)\cite{FDM1,FDM2,FDM3,FDM4,FDM5,FDM6,FDM7} remains an effective tool for solving perturbation equations in compact star spacetimes. In this section, we will employ FDM in the $r_*-t$ coordinate system to compute and analyze the dynamical behavior of axial perturbations in a charged uniform density star.

First of all, let's introduce the tortoise coordinates $y=\int_0^x\frac{dx}{\sqrt{f_x(x)h_x(x)}}$. By using the Fourier inverse transformation relation $\frac{\partial}{\partial \tau}\rightarrow-iw$, the wave equation Eq.(\ref{perturbation2}) is rewritten in the form in the time domain as follows
\bqn
\lb{FDM1}
\frac{\partial^2R_T}{\partial y^2}-\frac{\partial^2R_T}{\partial \tau^2}&=&V_TR_T+U_TR_V,\nb\\
\frac{\partial^2R_V}{\partial y^2}-\frac{\partial^2R_V}{\partial \tau^2}&=&V_VR_V+U_VR_T
\eqn
After the discretization process, above equation becomes
\bqn
\lb{FDM2}
\psi^{i+1}_j&=&-\psi^{i-1}_j+\frac{\Delta \tau^2}{\Delta y^2}\left(\psi^i_{j-1}+\psi^i_{j+1}\right)\nb\\
&&+\left(2-2\frac{\Delta \tau^2}{\Delta y^2}-\Delta \tau^2 V^T_j\right)\psi^i_j-\Delta \tau^2 U^T_j\phi^i_j ,\nb\\
\phi^{i+1}_j&=&-\phi^{i-1}_j+\frac{\Delta \tau^2}{\Delta y^2}\left(\phi^i_{j-1}+\phi^i_{j+1}\right)\nb\\
&&+\left(2-2\frac{\Delta \tau^2}{\Delta y^2}-\Delta \tau^2 V^V_j\right)\phi^i_j-\Delta t^2 U^V_j\psi^i_j ,\nb\\
\eqn
where $V^T_j\equiv V_T(j\Delta y)$, $U^T_j\equiv U_B(j\Delta y)$, $V^V_j\equiv V_V(j\Delta y)$, $U^V_j\equiv U_V(j\Delta y)$, $\psi_j^{i+1}\equiv R_T(\tau_0+i\Delta \tau,j\Delta y)$ and $\phi_j^{i+1}\equiv R_V(\tau_0+i\Delta \tau,j\Delta y)$.

According to Eq.(\ref{perturbation6}) and Eq.(\ref{perturbation7}), at the surface of star, the grid satisfy the relation
\bqn
\lb{FDM3}
\psi^i_{N_s}&=&\frac{\psi^i_{N_s-1}+\psi^i_{N_s+1}}{2},\nb\\
\phi^i_{N_s}&=&\frac{\phi^i_{N_s-1}+\phi^i_{N_s+1}}{2}.
\eqn
where $\Theta_N\equiv\Theta(N_s\Delta y)$.

By using the above difference scheme, we can employ the FDM for calculations. For convenience, we fix $M_S=0.25$ and $L=2$, and focus solely on studying the effect of parameter $Q_o$ on process of axial perturbation. The results are shown in Fig.\ref{Q000Fig}-Fig.\ref{Q800Fig}.
\begin{figure*}[!htbp]
\centering
\includegraphics[width=0.7\columnwidth]{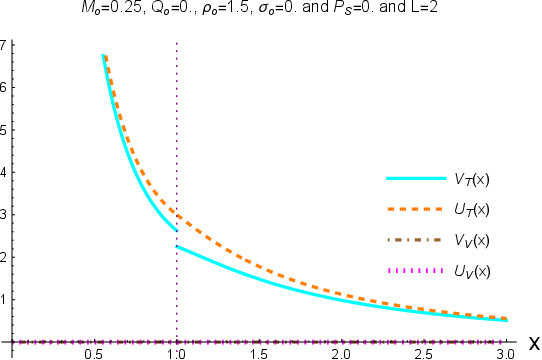}\includegraphics[width=0.7\columnwidth]{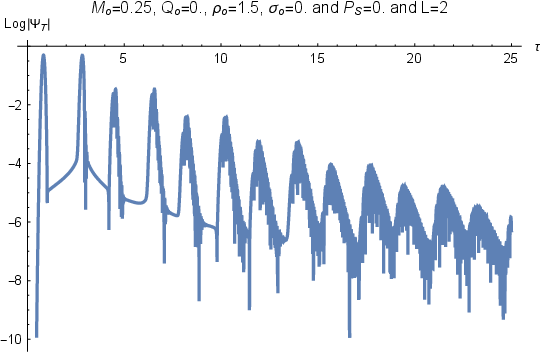}\includegraphics[width=0.7\columnwidth]{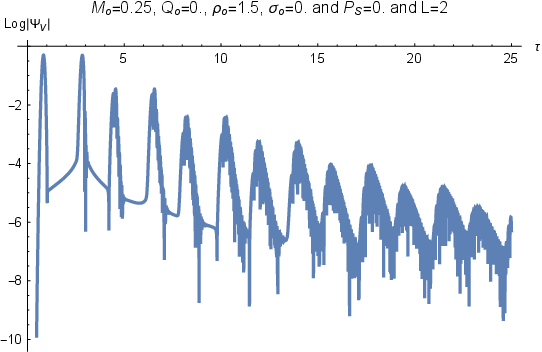}
\caption{The potential $V_T(x)$, $U_T(x)$, $V_V(x)$, $V_T(x)$ (the left graphic) and perturbation $\Psi_T$, $\Psi_V$ at surface with $Q=0$. The middle graphic ($\Psi_T$) and the right graphic ($\Psi_V$) show the dynamical behavior of axial perturbation in the $y-\tau$ coordinate system.
}
\lb{Q000Fig}
\end{figure*}

\begin{figure*}[!htbp]
\centering
\includegraphics[width=0.7\columnwidth]{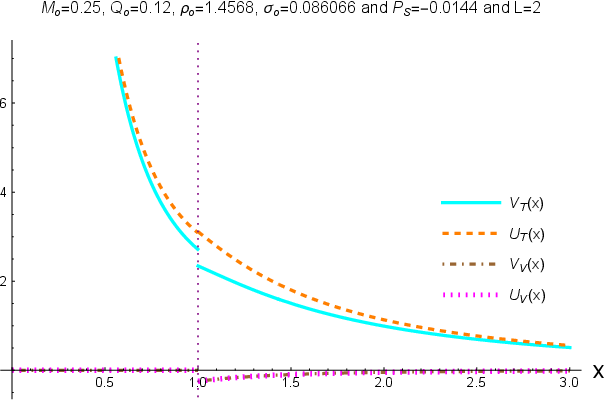}\includegraphics[width=0.7\columnwidth]{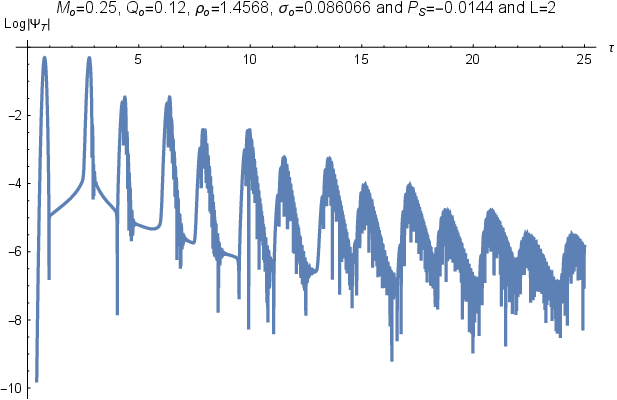}\includegraphics[width=0.7\columnwidth]{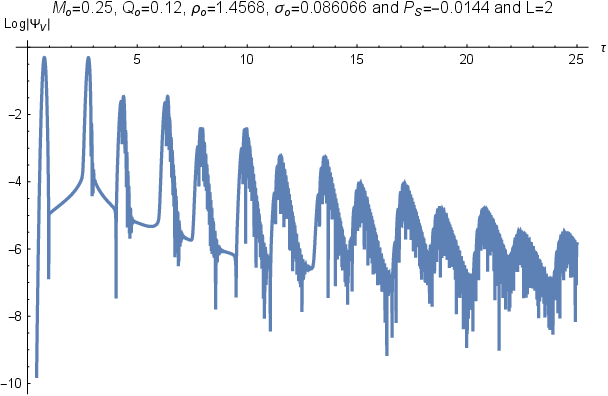}
\caption{The potential $V_T(x)$, $U_T(x)$, $V_V(x)$, $V_T(x)$ (the left graphic) and perturbation $\Psi_T$, $\Psi_V$ at surface with $Q=0.12$. The middle graphic ($\Psi_T$) and the right graphic ($\Psi_V$) show the dynamical behavior of axial perturbation in the $y-\tau$ coordinate system.
}
\lb{Q012Fig}
\end{figure*}

\begin{figure*}[!htbp]
\centering
\includegraphics[width=0.7\columnwidth]{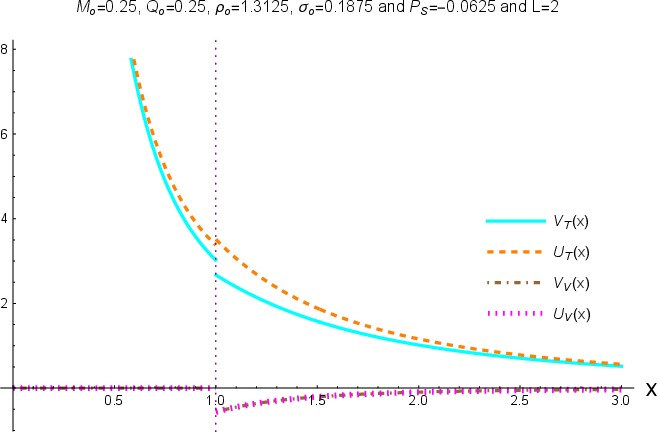}\includegraphics[width=0.7\columnwidth]{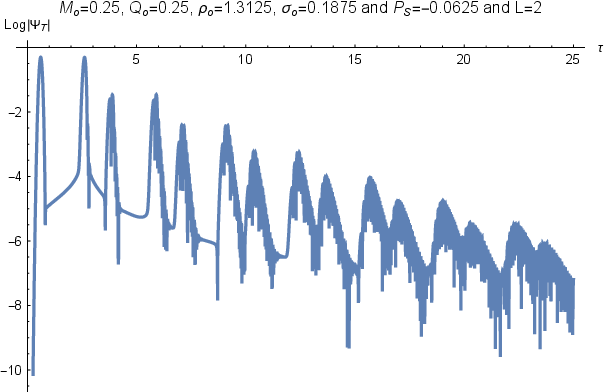}\includegraphics[width=0.7\columnwidth]{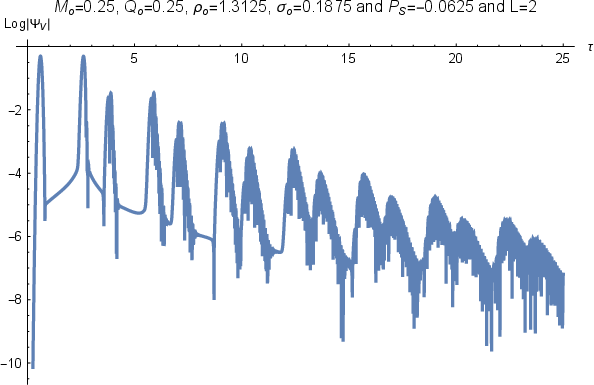}
\caption{The potential $V_T(x)$, $U_T(x)$, $V_V(x)$, $V_T(x)$ (the left graphic) and perturbation $\Psi_T$, $\Psi_V$ at surface with $Q=0.25$. The middle graphic ($\Psi_T$) and the right graphic ($\Psi_V$) show the dynamical behavior of axial perturbation in the $y-\tau$ coordinate system.
}
\lb{Q025Fig}
\end{figure*}

\begin{figure*}[!htbp]
\centering
\includegraphics[width=0.7\columnwidth]{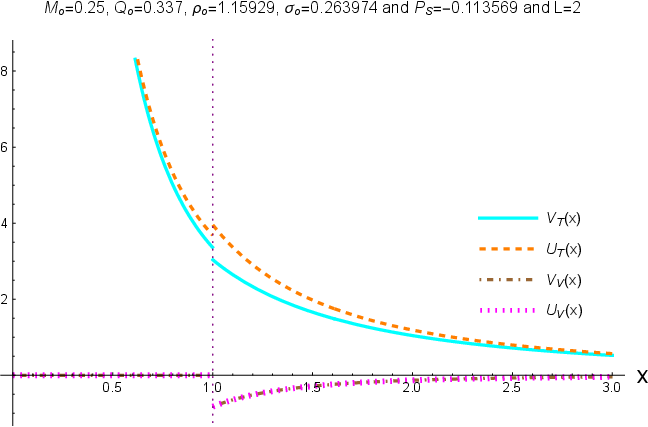}\includegraphics[width=0.7\columnwidth]{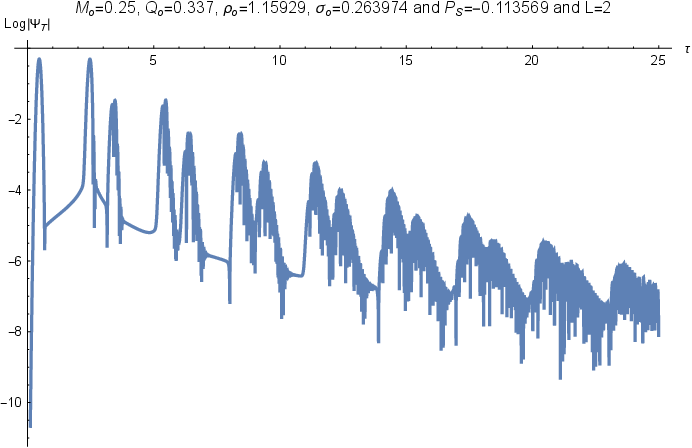}\includegraphics[width=0.7\columnwidth]{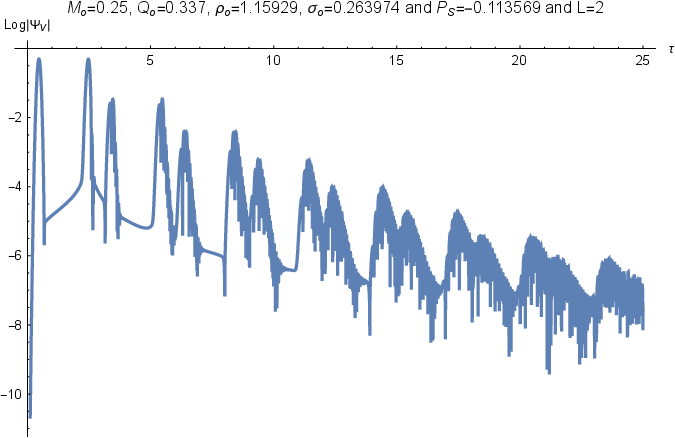}
\caption{The potential $V_T(x)$, $U_T(x)$, $V_V(x)$, $V_T(x)$ (the left graphic) and perturbation $\Psi_T$, $\Psi_V$ at surface with $Q=0.337$. The middle graphic ($\Psi_T$) and the right graphic ($\Psi_V$) show the dynamical behavior of axial perturbation in the $y-\tau$ coordinate system.
}
\lb{Q0337Fig}
\end{figure*}

\begin{figure*}[!htbp]
\centering
\includegraphics[width=0.7\columnwidth]{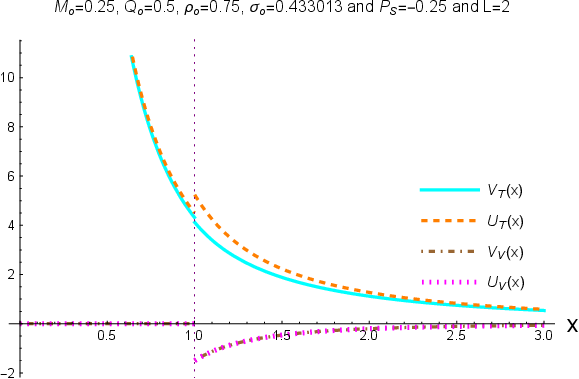}\includegraphics[width=0.7\columnwidth]{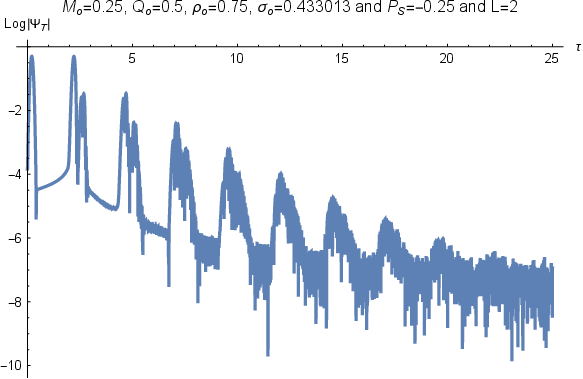}\includegraphics[width=0.7\columnwidth]{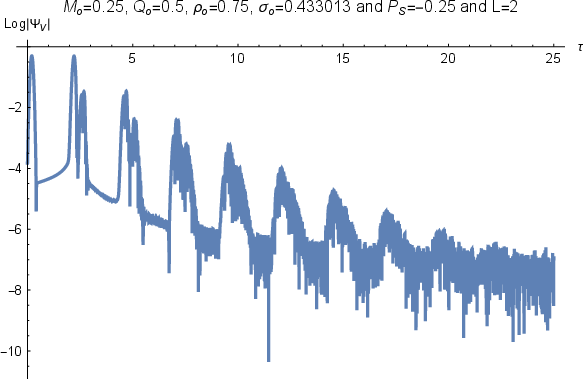}
\caption{The potential $V_T(x)$, $U_T(x)$, $V_V(x)$, $V_T(x)$ (the left graphic) and perturbation $\Psi_T$, $\Psi_V$ at surface with $Q=0.5$. The middle graphic ($\Psi_T$) and the right graphic ($\Psi_V$) show the dynamical behavior of axial perturbation in the $y-\tau$ coordinate system.
}
\lb{Q050Fig}
\end{figure*}

\begin{figure*}[!htbp]
\centering
\includegraphics[width=0.7\columnwidth]{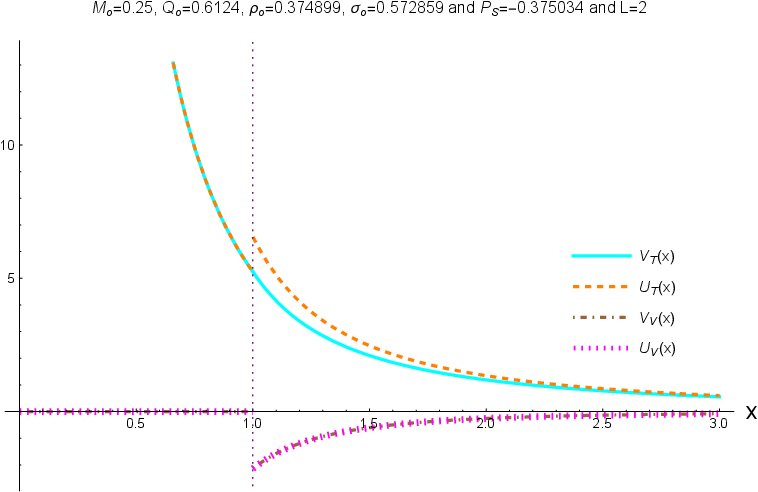}\includegraphics[width=0.7\columnwidth]{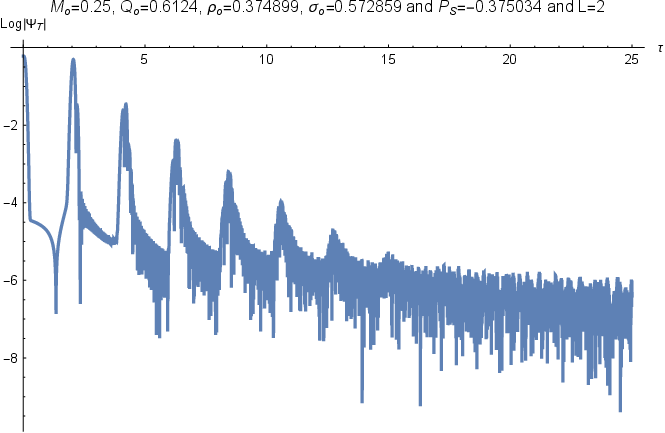}\includegraphics[width=0.7\columnwidth]{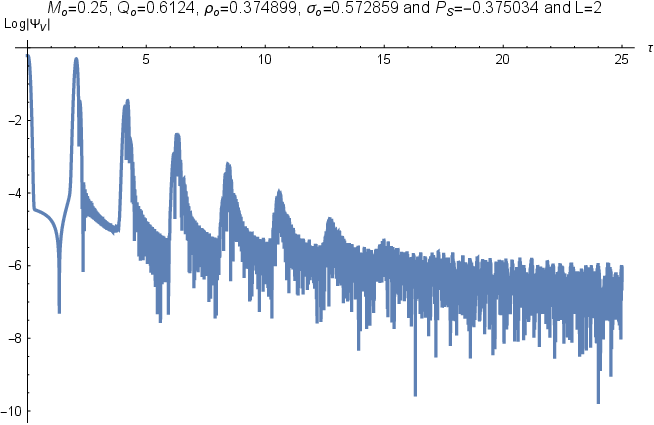}
\caption{The potential $V_T(x)$, $U_T(x)$, $V_V(x)$, $V_T(x)$ (the left graphic) and perturbation $\Psi_T$, $\Psi_V$ at surface with $Q=0.6124$. The middle graphic ($\Psi_T$) and the right graphic ($\Psi_V$) show the dynamical behavior of axial perturbation in the $y-\tau$ coordinate system.
}
\lb{Q06124Fig}
\end{figure*}

\begin{figure*}[!htbp]
\centering
\includegraphics[width=0.7\columnwidth]{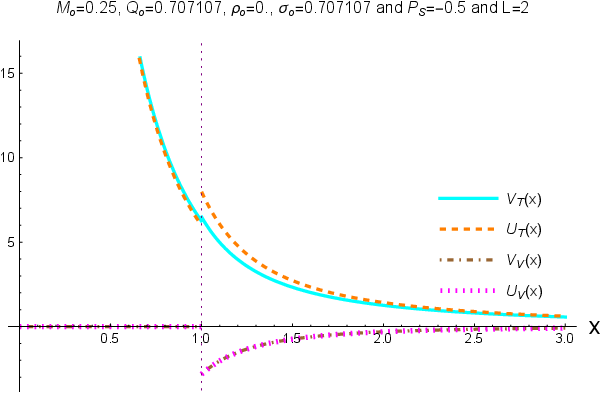}\includegraphics[width=0.7\columnwidth]{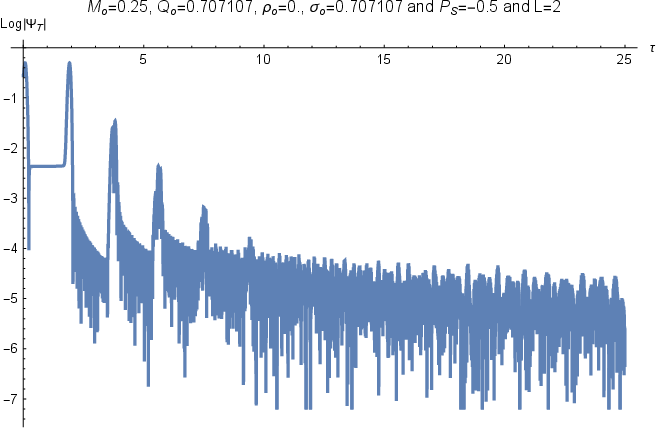}\includegraphics[width=0.7\columnwidth]{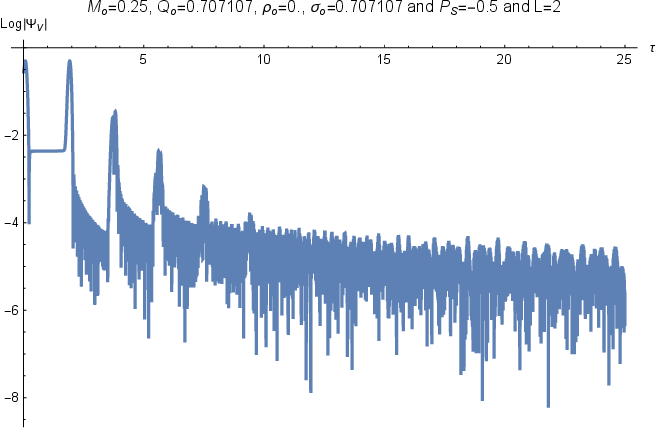}
\caption{The potential $V_T(x)$, $U_T(x)$, $V_V(x)$, $V_T(x)$ (the left graphic) and perturbation $\Psi_T$, $\Psi_V$ at surface with $Q=1/\sqrt{2}$. The middle graphic ($\Psi_T$) and the right graphic ($\Psi_V$) show the dynamical behavior of axial perturbation in the $y-\tau$ coordinate system.
}
\lb{Q0707Fig}
\end{figure*}

\begin{figure*}[!htbp]
\centering
\includegraphics[width=0.7\columnwidth]{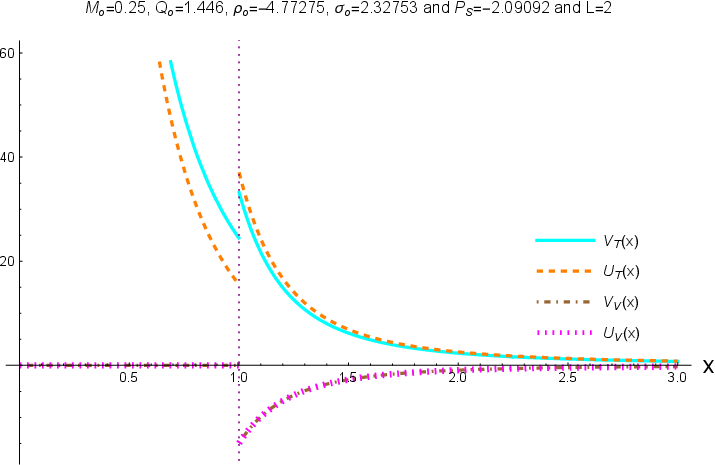}\includegraphics[width=0.7\columnwidth]{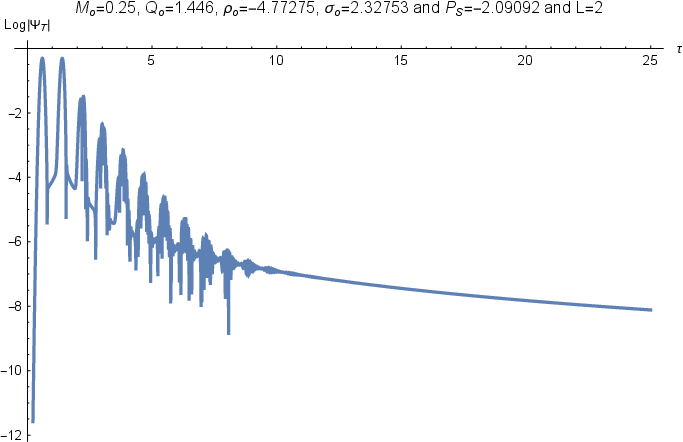}\includegraphics[width=0.7\columnwidth]{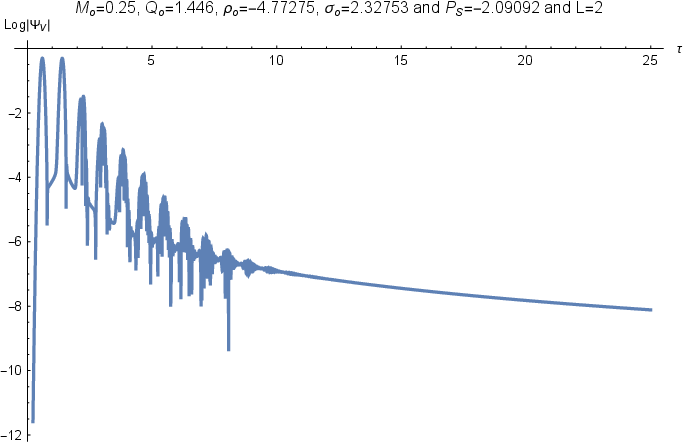}
\caption{The potential $V_T(x)$, $U_T(x)$, $V_V(x)$, $V_T(x)$ (the left graphic) and perturbation $\Psi_T$, $\Psi_V$ at surface with $Q=1.446$. The middle graphic ($\Psi_T$) and the right graphic ($\Psi_V$) show the dynamical behavior of axial perturbation in the $y-\tau$ coordinate system.
}
\lb{Q1446Fig}
\end{figure*}

\begin{figure*}[!htbp]
\centering
\includegraphics[width=0.7\columnwidth]{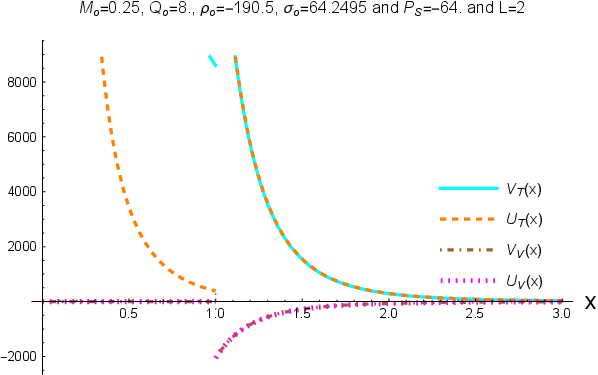}\includegraphics[width=0.7\columnwidth]{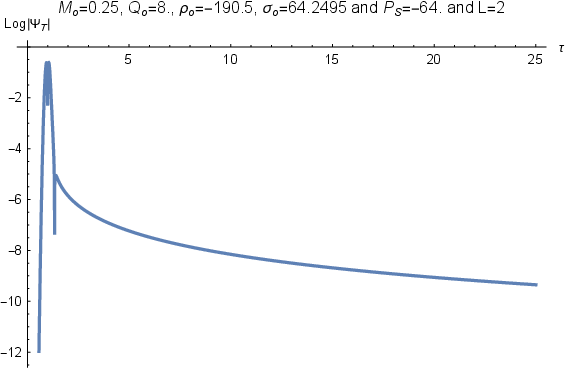}\includegraphics[width=0.7\columnwidth]{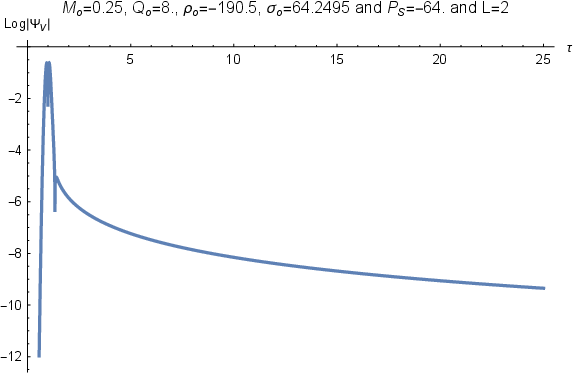}
\caption{The potential $V_T(x)$, $U_T(x)$, $V_V(x)$, $V_T(x)$ (the left graphic) and perturbation $\Psi_T$, $\Psi_V$ at surface with $Q=8$. The middle graphic ($\Psi_T$) and the right graphic ($\Psi_V$) show the dynamical behavior of axial perturbation in the $y-\tau$ coordinate system.
}
\lb{Q800Fig}
\end{figure*}

From above figures, it can be observed that these oscillation characteristics are influenced by echoes. As stated in \cite{echoTwo}, echoes originate from two sources: one is the potential well in the potential function, and the other is geometric surfaces with discontinuities, such as the stellar surface. Since the surface of a uniform density star always exhibits discontinuities or non-smooth features, this echo effect inevitably exists.

On the other hand, we also find that, due to the coupling of the two axial perturbation equations, the shapes and frequencies of the axial electromagnetic perturbation and the axial gravitational perturbation are remarkably similar, suggesting a synchronization phenomenon between them. Notably, since the coupling of $R_T$ and $R_V$ in the axial perturbation equations involves $A_x’$, this coupling is induced by the electrostatic charge $Q$.

From the above waveforms, we can further observe the following features. When $Q_o$ is relatively small, the perfect fluid inside the star consists of ordinary matter, satisfying $\rho > 0$. In this case, oscillations induced by echoes are clearly visible in the waveform and can persist for a relatively long time. Near $Q_o = Q_m \approx 0.707$, the matter density inside the star becomes very small. The echo-induced features in the oscillatory waveform gradually weaken, and after an initial oscillatory phase, a high-frequency oscillation emerges and persists for an extended duration. As $Q_o$ increases further, the star becomes composed of exotic matter, and the waveform gradually enters a tail-dominated regime. Finally, when $Q_o \gg M_o$, the quasinormal-mode oscillatory phase is progressively compressed until it nearly disappears. In this regime, the entire signal is dominated by a long-lasting tail, and the resulting waveform closely resembles an overdamped oscillation in damped systems.


\section{conclusion}
\renewcommand{\theequation}{6.\arabic{equation}} \setcounter{equation}{0}

In this paper, we treat the uniform density star as a spherically symmetric, perfectly conducting body, where all electric charges are uniformly distributed on the surface. We obtain analytical solution of the spacetime metric and find that due to the presence of surface charges, the pressure at the stellar surface is no longer continuous. We observe that when the fluid density inside the compact star is fixed, a small charge induces an outward repulsive force, requiring the fluid inside the star to provide an inward pressure to balance the electrostatic repulsion. However, when the charge becomes sufficiently large, the gravitational effect of the electric field dominates, generating an inward gravitational pull at the stellar center. To counteract this gravitational effect, the fluid inside the star must then provide an outward pressure.

Subsequently, we further investigated the quasinormal modes and dynamical behavior of perturbations in this spacetime under axial perturbations. We find that the axial perturbations of a charged uniform star are generally stable. However, the waveform is still strongly influenced by the charge. When the charge is relatively weak, the echo phenomenon in the waveform is pronounced. However, when the charge is strong, the synchronization between the axial Maxwell perturbation and the axial gravitational perturbation becomes more significant, and a high-frequency phase or a power-law tail phase may emerge.


According to the no-hair theorem of black holes, a black hole only exhibits three external characteristics: mass, charge, and angular momentum. Therefore, studying charged compact stars can help us better understand the role of charge in both classical and quantum gravity theories. On the other hand, compact stars with electromagnetic fields hold significant research value in astronomical observations. Hence, we plan to continue investigating related issues in the future.

\begin{acknowledgments}
The author thanks Elcio Abdalla, Eveling C. Ribeiro and Marcos R. Ribeiro for the helpful discussions and assistance.

The author acknowledges the funding support provided by the following agencies: FAPESQ-PB.
\end{acknowledgments}

\end{document}